# Engineering Space for Light via Transformation Optics

Alexander V. Kildishev and Vladimir M. Shalaev

*Birck Nanotechnology Center and School of Electrical and Computer Engineering, Purdue University, West Lafayette, IN 47907, USA*

Conceptual studies and numerical simulations are performed for imaging devices that transform a near-field pattern into magnified far-zone images and are based on high-order spatial transformation in cylindrical domains. A lens translating a near-field pattern from an almost circular input boundary onto a magnified far-field image at a flat output boundary is considered. The lens is made of a metamaterial with anisotropic permittivity and permeability both depending on a single 'scaling' parameter of the transformation. Open designs of the lens with a truncated body ('¾-body' and '¼-body' lenses) are suggested and analyzed. It is shown that the ideal full-lens and the '¾-body' lens produce identical images. Numerical simulations of '¼-body' designs indicate that further truncation of the lens could limit its performance. A light concentrator, "focusing" far-zone fields into a nanometer-scale area, is also considered.

Recently, increasing attention has been applied to creating an electromagnetic cloak of invisibility based on various schemes, including anomalous localized resonance,[1] dipolar scattering cancellation,[2] tunneling light transmittance,[3] sensors and active sources,[4] and transformation optics.[5-8] The latter approach was also employed for demonstrating cloak designs for the visible range.[9,10] However, transformation optics can go far beyond just cloaking because it allows one to control the path of light in unprecedented manner. By creating complex spatial distributions of dielectric permittivity $\varepsilon(r)$ and magnetic permeability $\mu(r)$, one can "curve" the optical space in any desired way and mold the flow of light.

In this Letter, we develop the conceptual basis for devices transforming near-field patterns into magnified (beyond the diffraction limit) far-zone images that can be detected with conventional optics. We also propose a "half-space" concentrator for light, which is different from the earlier proposed cylindrical concentrator [11], and represents, in a sense, a "reversed" magnifying lens. In contrast with [12] dealing with two concentric circular cylinder boundaries, new transformations connect a near-field pattern at an interior, almost circular cylinder, boundary with a far-field pattern at an exterior planar boundary.

In a general cylindrical coordinate system (CCS), $\rho$, $\phi$, and $z$, can be arranged by translating a $xy$-plane map ($x = x(\rho,\phi)$, $y = y(\rho,\phi)$) perpendicular to itself, and the resulting coordinate system forms families of concentric cylindrical surfaces. Given that $\vec{r} = \hat{x}x + \hat{y}y$, the scaling factors of a given CCS are equal to $s_\rho = |\vec{r}^{(\rho)}|$, $s_\phi = |\vec{r}^{(\phi)}|$, and $s_z = 1$, where $(\cdot)^{(\xi)} = \partial(\cdot)/\partial\xi$. It is convenient to choose a CCS with $s_\rho = s_\phi = s$. For the TM case, the curl of the magnetic field intensity, $\vec{H} = \hat{z}h$, is $s\nabla h \times \hat{z} = -h^{(\rho)}\hat{\phi} + h^{(\phi)}\hat{\rho}$ or $-\iota\omega s d_\rho \hat{\rho} - \iota\omega s d_\phi \hat{\phi}$; so the components of the electric displacement vector, $d_\rho$ and $d_\phi$, are equal to $d_\rho = (-\iota\omega s)^{-1} h^{(\phi)}$ and $d_\phi = (\iota\omega s)^{-1} h^{(\rho)}$. Transition to $h$ is given by $\left(\varepsilon_\phi^{-1} d_\phi\right)^{(\rho)} - \left(\varepsilon_\rho^{-1} d_\rho\right)^{(\phi)} = \iota s \omega \varepsilon_0 \mu_0 \mu h$, and the wave equation is defined by

$$\left(s^{-1}\varepsilon_\phi^{-1} h^{(\rho)}\right)^{(\rho)} + \left(s^{-1}\varepsilon_\rho^{-1} h^{(\phi)}\right)^{(\phi)} + \mu s k_0^2 h = 0. \quad (1)$$

In a virtual free-space ($\eta$, $\phi$, $z$)

$$\left(\tilde{s}^{-1} h^{(\eta)}\right)^{(\eta)} + \left(\tilde{s}^{-1} h^{(\phi)}\right)^{(\phi)} + \tilde{s} k_0^2 h = 0. \quad (2)$$

The simplest approach to transforming concentric cylindrical domains is built on a scaling transformation, $\rho = \rho(\eta)$, where parameter $\eta$ of the initial virtual domain is mapped onto a corresponding parameter $\rho$ in the physical world, keeping the other coordinate of the $xy$-plane $\phi$ intact. Then, Eq. (1) can be rearranged in the following manner using $\rho = \rho(\eta)$ and $f^{(\rho)} = \eta^{(\rho)} f^{(\eta)}$:

$$\left[\left(\frac{\eta^{(\rho)}\tilde{s}}{\varepsilon_\phi s}\right)\tilde{s}^{-1} h^{(\eta)}\right]^{(\eta)} + \left[\left(\frac{\tilde{s}}{\eta^{(\rho)} s \varepsilon_\rho}\right)\tilde{s}^{-1} h^{(\phi)}\right]^{(\phi)} + k_0^2\left(\frac{\mu s}{\eta^{(\rho)}\tilde{s}}\right)\tilde{s}h = 0.$$

To match the above equation for the field inside the lens with that of free-space in Eq. (2), the terms in parentheses should be equal to unity, i.e. the material of the lens should provide the following local properties:

$$\varepsilon_\phi = \eta^{(\rho)}\tilde{s}/s, \; \varepsilon_\rho = \tilde{s}/(s\eta^{(\rho)}), \; \mu = \varepsilon_\phi. \quad (3)$$

The aforementioned rules are valid for any device using a concentric mapping of cylindrical domains with either a linear or high-order scaling transform, $\rho(\eta)$, while preserving the common orthogonal pa-rameter $\phi$. For example, consider a CCS generated by the following *xy*-map,

$$x = \alpha\left[\frac{\pi}{2} - \tan^{-1}\left(\frac{\sinh\rho}{\cos\phi}\right)\right], \quad y = \frac{\alpha}{2}\ln\left(\frac{\cosh\rho + \sin\phi}{\cosh\rho - \sin\phi}\right), \quad (4)$$

with $s = \alpha\sqrt{2}/\xi_\rho$, for $0 \leq \rho < \infty$, and $-\pi/2 < \phi < \pi/2$; here $\alpha = 2x_0/\pi$ and $\xi_\rho = \sqrt{\cosh 2\rho + \cos 2\phi}$.

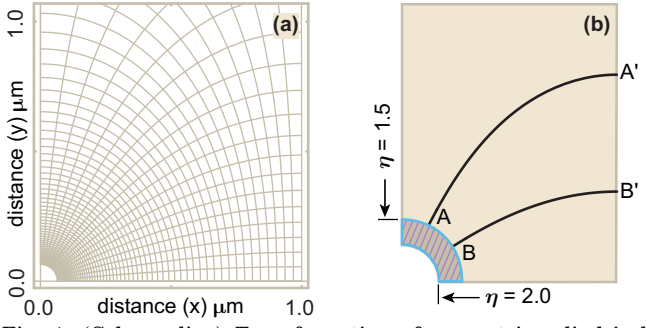

Fig. 1. (Color online) Transformation of concentric cylindrical domains. (a) A one-quarter xy-map generated by Eqs. (4) with $x_0 = 1\,\mu m$, $0 \leq \phi \leq \pi/2$, and $0 \leq \eta \leq 3$ (b) Mapping of virtual domain (hatched quarter-ring) onto physical domain (solid region), where the shared boundary is at $\rho = \eta = 2$. For example, points A and B from the curvilinear boundary at $\eta = 1.5$ are mapped onto the points A' and B' of the external boundary of the lens at $\rho = 0$.

For the ideal lens, a straightforward linear transform $\rho(\eta) = \tau(\eta - b)$ with $\tau = a(a-b)^{-1}$ gives $\eta = \tau^{-1}\rho + b$, and $\eta^{(\rho)} = \tau^{-1}$. Thus,

$$\varepsilon_\phi = \kappa/\tau, \quad \varepsilon_\rho = \kappa\tau, \quad \mu = \varepsilon_\phi, \quad (5)$$

where $\kappa = \xi_\rho/\xi_\eta$, $\xi_\eta = \sqrt{\cosh 2\eta + \cos 2\phi}$.

Fig. 1a depicts a one-quarter *xy*-map generated by Eqs. (4) with $x_0 = 1\,\mu m$, $0 \leq \phi \leq \pi/2$, and $0 \leq \rho \leq 3$. Fig 1b shows the mapping of a virtual domain (hatched quarter-ring) onto a physical domain (solid region) with the shared boundary at $\rho = \eta = 2$. Thus, points A and B from the virtual external boundary at $\eta = 1.5$ are mapped onto the points A' and B' of the physical external boundary of the lens at $\rho = 0$.

Calculation of the material properties requires the inverse transforms, which in this case are given by

$$\rho = \tfrac{1}{2}\ln\left(\frac{\cosh\widehat{y} + \cos\widehat{x}}{\cosh\widehat{y} - \cos\widehat{x}}\right), \quad \phi = \tan^{-1}\left(\frac{\sin\widehat{x}}{\sinh\widehat{y}}\right), \quad (6)$$

where the scaled coordinates $(\widehat{x}, \widehat{y})$ are defined as $\widehat{x} = x/\alpha$ and $\widehat{y} = y/\alpha$.

Finally, the anisotropic permittivity, $\vec{\varepsilon}$, is equal to

$$\vec{\varepsilon} = \frac{\kappa}{2}\left[(\tau + \tau^{-1})\mathbf{i} + \frac{\tau - \tau^{-1}}{\cosh 2\widehat{y} - \cos 2\widehat{x}}\mathbf{u}\right], \quad (7)$$

where $\mathbf{i}$ is a $2 \times 2$ idem factor; the components of the $2 \times 2$ symmetric matrix $\mathbf{u}$ are $u_{xy} = u_1$ and $u_{xx} = -u_{yy} = u_0$, with $u_1 = \sinh 2\widehat{y}\sin 2\widehat{x}$ and $u_0 = 1 - \cosh 2\widehat{y}\cos 2\widehat{x}$.

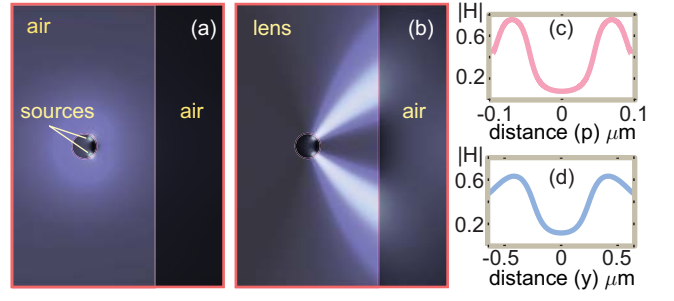

Fig. 2. (Color online) Test of near-to-far field projection. (a) The magnetic field intensity generated by two coherent test sources in air. (b) The magnetic field intensity generated by the pair of test sources inside and just outside the lens. A different color shade is taken inside the lens cross-section in (a) and (b). (c) The H-field magnitude generated by the sources along the curvilinear (input) surface of the lens, $\rho = \eta = 2$. (d) The H-field magnitude created by the sources at the flat (output) edge, $\rho = 0$.

Figure 2 shows a proof-of-concept test of the lens with smoothly changing local properties in accordance with Eqs. (5). The test sources in Fig. 2a are two cylindrical segments excited as magnetic line sources with $\lambda = 1.55\,\mu m$ at the virtual surface of a circular cylinder with a radius of 152 nm; the angular dimensions of the sources range from 10 to 40 degrees. In contrast with Fig.1a, where the sources give a combined (non-resolved) far-field pattern, in Fig. 2b the sources are resolved at the flat edge of the lens.

This closed design of the lens could limit its practical use. To provide better access to the input surface, a part of the lens can be removed. Figure 3abc shows the effect due to a symmetric truncation of the lens body. Two open designs of the lens are shown in Fig. 3ab, where the shaded sections indicate the lens cross-section in both panels. Fig. 3a depicts the H-field map inside a '¾-body' lens, while Fig. 3b shows the map for '¼-body' lens. The H-field magnitudes generated by the sources along the flat (output) boundary of the lenses are compared in Fig. 3c. The H-field at the image plane of the '¾-body' lens (dashed line) and the full-lens design (grey solid line) completely overlap. This good match is almost preserved in '½-body' lens (not shown), but the performance of the lens decreases with further truncation. For example, Fig 4c shows additional artifacts and decreasing intensity at the image boundary.

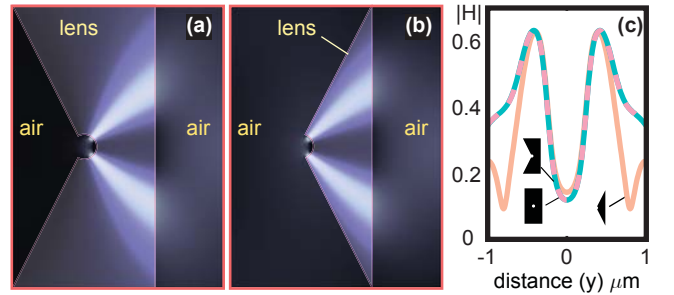

Fig. 3. (Color online) Open designs of the lens. (a) The magnetic field map inside a '¾-body' lens. (b) The field map inside a '¼-body' lens. Shaded areas indicate the lens cross-section in (a) and (b). (c) The H-field generated by the sources along the flat surface of either the closed or open lens designs.

Finally, a similar approach is taken to make a light-concentrating device, where for the ideal case another common boundary, $\rho = \eta = 0$, is used along with the simple transformation $\rho = \tau \eta$, $\tau = l/b$, which results in mapping a virtual boundary $\eta = b$ onto a physical boundary inside the lens ($\rho = l$). The transform gives the same formal properties as (5).

Fig. 4 compares the performance of open designs of light-concentrating devices, shown in panels (b), (c), and (d) versus the full-scale concentrator shown in panel (a). All the panels show the time-averaged energy density in logarithmic scale, while the concentrators are illuminated by 750-nm plane wave propagating from right to left. The geometrical and transform parameters of all the concept devices in Fig. 4 are $x_0 = 1\ \mu m$, $l = 200$ and $b = 0.01$. Panels (a) and (b) depict designs with identical performance; the focusing properties of the open design in (b) are not affected by the truncation of the device body. In contrast with the designs of Fig. 4ab, the open designs of Fig. 4cd demonstrate visible distortion and undesired localization of energy density due to additional interfaces interrupting the smooth transformation of critical spatial modes.

In summary, this Letter deals with the proof-of-concept numerical simulations of near-to-far field magnified imaging and light-concentrating devices built on high-order spatial transformations in general cylindrical domains. The primary results are (i) a lens translating a near-field pattern generated at an almost circular input boundary onto a magnified far-field pattern at a flat output boundary, and (ii) a light concentrator focusing the energy of a plane-wave input into a sub-wavelength spot. The lens and the concentrator are modeled out of a metamaterial with anisotropic permittivity and permeability both depending only on a single parameter ($\rho$) of an orthogonal coordinate system that defines the transformation. Two open designs of the lens with truncated body ('¾-body' and '¼-body' lenses) are studied and compared with a full, closed design. It is shown that the ideal full-lens design and the truncated '¾-body' lens produce identical images. Similarly, the '¾-body' light concentrator provides performance comparable with the full body ideal design. Numerical studies of the '¼-body' lens and concentrator indicate that further truncation of the device could substantially limit its functionality.

The authors are grateful to Prof. Narimanov for useful discussions. This work was supported in part by ARO grant W911NF-04-1-0350 and by ARO-MURI award 50342-PH-MUR.

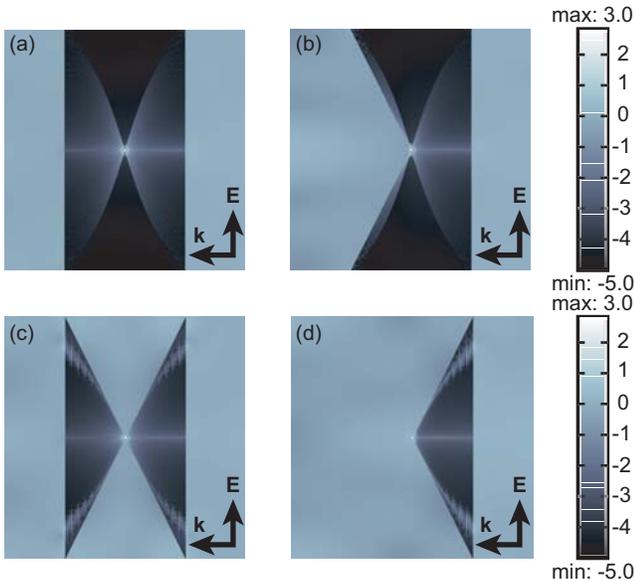

Fig. 4. (Color online) Comparison the full design of the light-concentrator (a) with an open '¾-body' design (b). (c) and (d) '½-body' and '¼-body' designs of the concentrator. All the panels show the time-averaged energy density in logarithmic scale for TM polarized 750-nm plane wave propagating from right to left.